\documentstyle[12pt,aasms]{article}
\def\etal{{\it et al. }}

\begin{document}

\title{Ellipticals with Kinematically--Distinct Cores:\\
WFPC2 Imaging of Globular Clusters\altaffilmark{1}}

\author{Duncan A. Forbes}
\affil{Lick Observatory, University of California, Santa Cruz, CA 95064}
\affil{Electronic mail: forbes@lick.ucsc.edu}

\author{Marijn Franx} 
\affil{Kapteyn Institute, University of Groningen, P.O. Box 800,
9700 AV Groningen, The Netherlands}
\affil{Electronic mail: franx@astro.rug.nl}

\author{Garth D. Illingworth}
\affil{Lick Observatory, University of California, Santa Cruz, CA
95064}
\affil{Electronic mail: gdi@lick.ucsc.edu}

\author{C. M. Carollo}
\affil{Leiden Observatory, 2300 RA Leiden, The Netherlands}
\affil{Electronic mail: carollo@strw.leidenuniv.nl}

\altaffiltext{1}{Based on observations with the NASA/ESA {\it Hubble
Space Telescope}, obtained at the Space Telescope Science Institute,
which is operated by AURA, Inc., under NASA contract NAS 5--26555}

\begin{abstract}

New globular clusters may form in the merger of two galaxies.
Perhaps the best examples of merger remnants are the set of
ellipticals with kinematically--distinct cores. 
Here we present {\it HST} WFPC2 imaging of 14 
kinematically--distinct core ellipticals to examine their globular
cluster systems. In particular, we probe the galaxy central regions,
for which we might expect to see the strongest signatures of some
formation and destruction processes. These data substantially increase
the number of extragalactic globular cluster systems studied to date. 
We have developed a method for galaxy subtraction and
selection of globular clusters which results in about two hundred
globulars per galaxy to a limiting magnitude of V $\sim$
25. Simulations of artifical globulars are also described. 

We find that the globular cluster luminosity, 
and color, vary only weakly, if at all, 
with galactocentric distance. The mean color of globular
clusters are constant with globular cluster magnitude. 
Several clear trends are also present. 
Firstly, globular cluster colors are bluer (more metal
poor by $\sim$ 0.5 dex) than the underlying galaxy starlight at any
given galactocentric distance. Second, we find a correlation
over roughly ten magnitudes between the mean globular
cluster metallicity and parent galaxy luminosity of the form Z
$\propto$ L$^{0.4}$. This relationship includes dwarf ellipticals,
spiral galaxy bulges and giant ellipticals. Third, 
we find that globular cluster surface density distribution 
can be described by a core model, for which the core radius correlates
with galaxy luminosity. Lastly, for the sample as a whole, the globular cluster
systems are closely aligned with the galaxy major axis and are
slightly rounder than the galaxy itself, although their are some notable
exceptions. 

Our results favor scenarios in which ellipticals form from massive,
gas rich progenitors at early epochs. Detailed simulations of the
formation of globular cluster systems would be valuable to draw firmer
conclusions. 

\end{abstract}


\section{Introduction}

Globular clusters, as well as being interesting in their own right,
provide 
important clues to the formation 
and evolution of their parent galaxy (see for example Harris 1991 and
references 
therein). In particular, 
the merger of two spiral galaxies may lead to the formation of an
elliptical galaxy and create an 
additional population of globular clusters (GCs) in the process 
(Schweizer 1987). This potentially 
overcomes the objection raised by van den Bergh (1990) that
ellipticals have many
more GCs, per unit light, than expected from simply 
combining the observed GC population of two spirals. 
This idea was expanded upon by Ashman \& 
Zepf (1992). In their model, new GCs are formed in 
massive gas clouds from the progenitor spirals. These gas clouds are
of higher metallicity and are more centrally concentrated  
than the original GCs 
belonging to the spirals. Thus an observational consequence
of such a model is a radial metallicity 
gradient and at least two peaks in the metallicity distribution of the
resultant GC population. 

Kinematically--distinct core (KDC) ellipticals have central regions that
rotate rapidly, and often in an opposite 
direction, compared 
to the stars in the outer parts of the galaxy. They are
found in about 
1/3 of all ellipticals 
and are generally thought to be the result of a merger 
(Illingworth \& Franx 1989). 
This merger may involve the accretion of a small secondary 
(Balcells \& Quinn 1990), although this model has difficulty 
explaining the metallicity (Bender \& Surma 1992) 
and surface brightness (Forbes, Franx \& Illingworth 1995) 
profiles of some KDC galaxies. A more plausible alternative involves
the merger of gas--rich galaxies in which gas dissipates into the
central regions and subsequent star formation leads to
a KDC. Hernquist \&
Barnes (1991) have studied this for the case of two near equal mass
disk galaxies. 
Many KDC galaxies show evidence 
for small scale dust lanes and rings (Forbes, 
Franx \& Illingworth 1995) further supporting the
idea that these 
galaxies have accreted 
material. Thus KDC galaxies offer an ideal sample in which to
study 
their GC population and 
critically examine the hypothesis of merger--induced GC formation. 
In addition, such ellipticals cover a range of 
environments (which may affect the relative number of GCs and
metallicity; Harris 1991) 
and X--ray luminosities (which could 
contribute to GC formation via a cooling flow; Fabian, Nulsen \&
Canizares 1984). The results for KDC
ellipticals may also be applicable to other elliptical galaxies, as
our WFPC1 study indicated that the photometric properties of KDC
ellipticals follow the same scaling relations as `kinematically normal'
ellipticals (Forbes, Franx \& Illingworth 1995).

Previous studies of GC systems in kinematically--distinct core galaxies
have included NGC 4278 (Harris \& van
den Bergh 1981), 
NGC 4365 (Harris \etal 
1991), NGC 4406 (Hanes 1977), NGC 4472 (Harris 1986; Couture \etal
1991) and the work of
Ajhar \etal (1994)  
who studied NGC 4365,  
NGC 4406, NGC 4472 and NGC 4552. However these ground--based studies,
and those of others, 
are limited to relatively large galactocentric 
radii by seeing and blending effects, whereas we might expect the
inner regions to have the strongest signatures of some 
formation and destruction mechanisms. In a review, Harris (1993)
stated that ``...essentially nothing is known about the inner globular
clusters in most galaxies, and much new observational work could be
done in this direction.''

Here we present F555W (V) and F814W (I) Wide Field and Planetary
Camera 2 (WFPC2) images of 14 ellipticals with 
well--established KDCs. 
The superior resolution of {\it HST} makes it possible 
to probe GC properties within one effective radius
of the parent galaxy. To date only two other ellipticals, NGC 4881 (Baum
\etal 1995) and M87 (Elson \& Santiago 1996; Whitmore \etal 1995), have 
published WFPC2 data on their GC system. 
We examine the magnitude, color and spatial
distributions of the GC system in the central few kpc 
of each galaxy.    
Results for the parent galaxies themselves, and implications for the
formation mechanisms of KDCs, will be presented 
elsewhere (Carollo \etal 1996). We assume 
h = H$_o$/100 km s$^{-1}$ Mpc$^{-1}$ throughout the paper, although 
in a future paper we will discuss the use of our 
GC luminosity functions for the derivation of the Hubble constant. 

\section{Observations and Initial Data Reduction}

A total of 14 kinematically--distinct core galaxies were observed with
{\it HST's} WFPC2 between 1994 April and November. 
These galaxies represent most of the well--established cases
known to date. Their properties are summarized in Table 1. 
Data for each
galaxy consist of a pair of 500s F555W images and a pair of 230s F814W
images. 
The nucleus of each galaxy is located near the center of the PC CCD.  All
4 CCDs are 800 $\times$ 800 pixels, with a scale of 0.046$^{''}$/pix
for the PC and 0.097$^{''}$/pix for the WFC CCD.  

The standard STScI pipeline software was used for the initial data reduction. 
Before combining each pair of images, we needed to check the
relative alignment. We find that 
each pair of F555W and F814W images are aligned to within $<$ 0.1 WFC
pixel 
($<$ 0.01$^{''}$), and so can be 
combined without any shifting. Each pair of images is combined using
the 
STSDAS task {\it crrej} which has 
been modified by J. Biretta for WFPC2 images. The raw frames contain
a large 
number of cosmic rays, hot pixels 
and cold pixels. Hot pixels are created at a rate of hundreds per month 
(Holtzman \etal 1995a) 
and tend to be single pixel events, 
whereas cold pixels (CCD traps) are more stable with time. The
combining 
task does a very good job of 
removing cosmic rays and to some extent hot pixels (see discussion
below). 
Using lists of CCD traps kindly 
supplied by the WFPC group at STScI, we have interpolated over all 
such pixels. 
We have also trimmed 
each image to exclude the `obscured region' near the pyramid edges, 
yielding 751 $\times$ 751 pixel images. The final trimmed images are
missing $<$1\% of the effective area because the WFPC2 field-of-view
actually overlaps each CCD by about 40 pixels. 

We then made a first order attempt to remove the galaxy
background. For the WFC 
images there is a slight background 
gradient which is strongest towards the direction of the pyramid
center 
(which is the direction of the galaxy 
center). This gradient is largely removed by a bi--cubic spline fit. 
For the PC image, we fit the galaxy 
isophotes using the STSDAS task {\it ellipse} (see Forbes \& Thomson
1992). 
By subtracting these models 
we are left with globular clusters, foreground
stars, 
small faint galaxies and 
possibly dust features at the center of the galaxy. An example of a 
background--subtracted F555W image for 
NGC 1427 is shown in Fig. 1 (Plate **). 
Most of the objects are globular
clusters, 
and similar numbers appear on each image
(for the PC image, the increased cluster density roughly compensates for the 
reduced areal coverage). 

\subsection{Globular Cluster Selection and Measurement} 

The selection and photometry of GCs is carried out using the IRAF 
version of {\it daophot} (Stetson 1987). 
Selection of objects is based on the F555W image, which has a higher
S/N than the F814W image for the objects of interest. The same 
{\it daofind} parameters are used for each galaxy. In particular, we
use roundness parameters --0.5 to 0.5 and a 
threshold in counts of 4~DN (data number) for the PC and 5~DN for WFC images. 
We have adopted a fairly tight shape constraint so that the detected
objects are
not too distorted by noise or nearby dust features, which may also
affect their magnitudes and colors. 
The disadvantage of this approach is that a few faint sources, that
are probably real GCs, will not be detected by {\it daophot}, although 
measurement of these objects would give spurious results and so they
would be rejected from the subsequent analysis anyway. 
Tests on actual WFC images indicates that to V = 25, $\sim$10\% of
detected sources are more elongated that our roundness
criterion. Visual inspection of these sources suggests that about half
(5\%) are clearly not GCs, i.e. they are noise, diffraction spikes
of stars or galaxies. Thus we are excluding less than 5\% of GCs with our
conservative roundness criterion. 
A similar situation is present for PC images in which there
is no dust. For the dusty galaxies, {\it daophot} detects many sources
more elongated than our roundness constraint. Here it is more
difficult to make conclusive statements, however a visual inspection
of these images suggests that {\it daophot} is only missing a small
number ($\sim$ 5 per galaxy) of possible GCs to V = 25. The colors of
GCs located at the edges of dust patches is discussed in section 4.4. 

For the PC image, the
limiting magnitude for point sources 
is about 0.3$^m$ brighter than for the WFC images 
(Burrows \etal 1993). To compensate for this we chose a slightly lower
count threshold for the detection of objects, i.e. 4~DN (data number) 
instead of
5~DN. In order to confirm that all four CCDs have 
similar detection characteristics, we have combined the objects
detected by {\it daophot} for NGC 4365 and NGC 5982 (the only two
galaxies in the sample with no obvious dust in the PC or WFC frames). 
We find that each CCD has a similar detection fraction of  
objects to V = 25. 

We also tested an alternative method of selecting objects
independently on both the F814W and F555W images, and including 
only  those objects in common to give an object list. This method
gave significantly fewer objects
at faint magnitudes 
because the F555W image has a higher effective S/N
for GCs than the F814W image. 
We believe that the additional faint objects (found using the
selection in the F555W image only method) are real detections, as
their mean color is indistinguishable from that of the brighter sources. 

Photometry of the detected objects is carried out for both 
the F814W and F555W images using the coordinates of the objects
selected from the F555W image. Any 
residual background counts remaining from the 
background fits are removed by 
doing `sky' subtraction in an annulus around each object. 
We have chosen to measure magnitudes in a small aperture and make
aperture corrections in order to reduce the noise from the background
in very large apertures (the WFPC2
PSF actually contains some light out to 30 WFC pixels). 
The optimum aperture for stars has
been determined by several workers to be 2--3 WFC pixels radius (Whitmore 1995). 
Our tests of the
scatter in V--I color of actual GCs indicates that 
to V = 25 the dispersions between 
2 and 3 pixel apertures deviate by $<$5\% 
(in the sense that the 2 pixel aperture has a 
smaller dispersion). However with the 2 pixel apertures, there are
several issues that may cause a systematic bias in the final
photometry, such as centering,
small area calculations and the largely unknown variation of aperture
corrections with position on the CCD. We have taken what we believe
to the conservative approach and used 3 pixel radius apertures. 
These 
aperture magnitudes are converted to total magnitudes using the
encircled 
energy tables of Holtzman \etal 
(1995a), i.e. for F555W the correction is 0.23 and 0.16 mags 
for the PC and WFC CCDs  
respectively, and for F814W, 
0.32 and 0.16 mags. These aperture corrections are within 0.05 mags of
those used by Whitmore \etal (1995) based on measurements of isolated
stars. 
The geometric distortion between F555W and F814W is
negligible. In order to convert WFPC2 flight magnitudes to a
ground--based standard photometric system we use (Holtzman \etal 1995b):\\

V = 21.724 -- 2.5~log(DN/s) -- 0.052~(V--I) + 0.027~(V--I)$^2$
\hspace{5cm} (1)\\

I = 20.840 -- 2.5~log(DN/s) -- 0.063~(V--I) + 0.025~(V--I)$^2$
\hspace{5cm} (2)\\

\noindent
These equations are for a gain setting of 15, which applies to all but
three of our galaxies (we added 0.745$^m$ to the 
zeropoints for gain setting 7). 
Assuming (V--I) = (F555W mag. -- F814W mag.) in the above relations 
introduces a rms error of $<$ 1.5\% in the final magnitude
determination given the range of GC colors. 
Finally we have corrected the magnitudes for Galactic extinction,
listed in Table 1. We have restricted the detected magnitude range to
19 $<$ V $<$ 26, as there appear to be no GCs brighter than V $\sim$
20 (brighter sources appear to be stars) and fainter than 
V $\sim$ 25 (this is our effective limiting magnitude and 
will be discussed further in the next section).

For each selected object we have fit the profile with a 
Gaussian and measured its FWHM size, determined R.A. and
Dec. positions using the STSDAS 
task {\it metric}, calculated position angle on the sky, and the
radial distance from the galaxy center. Thus we now have a list of  
position, V magnitude and its error, I magnitude and its error, V--I
color and its error from a quadrature sum of the photometric errors, 
size, position angle and 
galactocentric distance for each object. 

At this point the list may include hot pixels,  
foreground stars, faint galaxies as well as globular clusters. 
Hot pixels can be effectively excluded by comparing the positions of
detected objects (in pixel coordinates) to the location of known
hot pixels shortly after the galaxy observation date. These known hot pixels 
should be a superset of the hot pixels for our actual data. 
The exception might be NGC 1700 which was observed in 1994 April 17 before the 
April 23 thermal cycling (we rely on 
object sizes and colors to check the reality of selected objects for this
galaxy). We have checked to see if any 
hot pixels lie within 3 pixels of our selected
objects, and typically find only half a dozen coincidences per galaxy. 
These objects are then removed from the list.

Elson \& Santiago (1996) estimate that a typical high galactic
latitude field contains about 2 foreground stars per WFC image in the
magnitude range 20 $<$ V $<$ 26. The PC image contains a smaller
number. Thus contamination by foreground stars is very small. 
Furthermore, we have removed all objects with FWHM size $\ge$ 3.0
pixels (typically 5 per image).  
This is about the size expected for the large Galactic globular
cluster Omega Cen if it were placed at Virgo distance (h = 0.75) and
convolved with {\it HST's} PSF (Grillmair 1995). Thus for 
intrinsically smaller globulars or ellipticals more distant than
Virgo, the expected
globular size is $<$ 3 PC pixels. 
Visual inspection reveals that most objects with sizes $>$ 3 pixels are
probably background galaxies. 
Studies of random high galactic latitude WFPC2 fields 
suggest that the number of small (FWHM $<$ 3 pixels) 
galaxies to I = 22 (V $\sim$ 23.5) 
is perhaps only one or two per WFC CCD image (Forbes \etal 1994; Phillips
\etal 1995). Galaxies fainter than V = 24 are generally not detected
in these images.

\subsection{Errors and Completeness Tests}

Photometric errors and sample completeness have been estimated in a
manner similar to that described by Secker \& Harris (1993).  
Using the IRAF task
{\it psf} we created an artificial object from the average shape
of $\sim$20 real GCs. This was done for both filters in a 
WFC image.  
We then populated a relatively empty image (the few 
existing GCs are interpolated across) with artificial
GCs. The noise characteristics of this image  
are fairly representative for all galaxies. 
The objects were placed at random locations, using the task {\it
addstar}, in 1 mag bins to cover the range 21 $<$ V $<$ 25.5. 
Objects in the I images were generated at the same position as in the
V images with an input color of V--I = 1.0. 
Several hundred artificial GCs were 
created, but no more
than 150 per image to ensure that blending of objects did not
occur (of the total available area, 150 GCs covers less than 1\%).  
Even for the richest GC systems in our sample, there is no
evidence of blending. The next step is to find and measure magnitudes
of these
artificial GCs. This is done using {\it daophot} with the same
selection and photometry parameters as for the real data. 

In Fig. 2 we plot the difference in the actual input magnitude of the
artificial GC and its measured magnitude as a function of input 
magnitude. This gives us an estimate of the measurement, or
photometric error, for our sample. The photometric error is fairly
symmetric about zero, showing that random errors dominate over any
systematic bias. Taking the absolute value of the error, we 
fit an exponential of the form\\

pe(m) = exp [ a (m - b)] ~~~~~~~~~~~~~~~~~~~~~~~~~~~~~~~~~~~~~~~~~~~~~~~~~~~~~~~~~~~~~~~~~~(3)\\

\noindent
to describe the photometric error. Here $pe$ is the absolute value
of the input minus output magnitude, $m$ is the GC magnitude, $a$ and
$b$ are coefficients. For the V band data we find that $a = 0.83$ and $b = 26.84$,
and for the I band data $a = 1.07$ and $b = 23.99$. This gives a photometric 
error of 0.1$^m$
at V $\sim$ 24 and I $\sim$ 22. We can compare these error functions
to the measurement errors determined directly by {\it daophot} for the
real GCs. Reassuringly, the variation of the measurement 
error with magnitude is
very similar to that determined from our tests described above. This 
gives us confidence in the magnitude and color errors given by {\it
daophot}. 

Although we will not discuss GC luminosity functions, it is still of
interest to determine the completeness function for our data, i.e. 
quantify the ability of {\it
daophot} to detect GCs as a function of magnitude. 
Here we use the
same input lists of artificial GCs as above, but now consider whether each
added GC was recovered by {\it daophot}. The ratio of the number of
input objects to those detected gives the sample completeness. 
The V completeness function ($cf$) for a WFC image is shown in Fig. 3a. 
We find that our sample is essentially
100\% complete to V $\sim$ 24,  and that our detection limit is 
V $\sim$ 25. The $cf$ is fit over the
range of interest with two exponentials. The first, $cf(m) = a 
$exp$[ b / (m - c)]$ is fit to magnitudes brighter than the 50\%
completeness level and $cf(m) = a / $exp$[b (m - c)]$ to
magnitudes fainter than the 50\% level, where $a,b$ and $c$ are
coefficients.  At the 50\% completeness 
level these functions agree within  $\pm 0.03^m$. To first order we
would expect all four CCDs to have a similar completeness function 
since the detection rate as a function of magnitude is similar for each
CCD.

\section{Results}

\subsection{Simulations}

Before presenting the real data, we need to check whether our detection
and photometry procedure has introduced any obvious bias. As 
mentioned earlier, the variation of {\it daophot} computed 
photometric errors with total magnitude is consistent with that from
the simulations. It is possible that our procedure introduces a
detection or photometric bias that depends on distance from the galaxy
center. 
In figures 3b and 3c we show
the radial V magnitude and V--I color profiles using the same
detection and photometry methods as for the real data. 
The V magnitudes of the simulated GCs were randomly distributed
between V = 21 and V = 26, while the I magnitudes
were set to be I = V -- 1.0. 
We find that the simulated GCs do {\it not} show a trend with galactocentric
radius in either magnitude or color.  
Thus any statistically significant radial
gradients seen in the actual WFPC2 data are likely to be real. 

\subsection{Globular Clusters}

Color--magnitude diagrams for the sample galaxies are presented in
Fig. 4. These show that most 
galaxies reveal a relatively tight concentration of points
around V--I$\sim$ 1, and that this average color is relatively constant with
magnitude. The dispersion in color increases at faint magnitudes. 
At the detection limit we would expect the average color to
become redder as we are biased against detecting blue objects but this
is not seen due to our conservative magnitude cutoff. 
The raw (i.e. not corrected for incompleteness) globular cluster
luminosity functions (GCLFs) are shown in Fig. 5. A variety of shapes are
seen. For some galaxies the expected apparent magnitude peak in the
GCLF, if it is universal, is within our completeness limit (e.g. NGC
4278) while for others it is well beyond (e.g. NGC 1700). 
Globular cluster luminosity functions corrected for incompleteness 
will be discussed in a future paper. 
The distribution of globular cluster colors for
our sample is shown in Fig. 6. These histograms show 
that there is a fairly small range in the mean GC color from galaxy to
galaxy.

At this stage, we have decided to apply an additional selection
criterion based on GC color. For the subsequent results and analysis
given below, we will restrict the data set for each galaxy to GCs with
colors within 3$\sigma$ of the mean. 
This should remove any objects with 
deviant colors that are 
due to undetected cosmic rays or noise spikes. Such a selection
process removes typically only a small fraction of objects from the list.
Nevertheless, we have checked the rejected objects to ensure that we
are not excluding the most interesting candidates. Objects with extreme
red or blue colors do {\it not} lie preferentially at small or large
galactocentric distances, so we are not for example removing centrally
located blue GCs. The objects with extreme colors tend also to be the
faintest sources (see Fig. 4) 
suggesting that photometric errors play a large role 
in determining their deviant colors. 
We are confident that our final 
list does 
not contain any hot pixels or resolved galaxies
and that the contribution from cosmic rays,  
foreground stars and unresolved galaxies is negligible. 
%Real globular clusters should dominate our lists. 

After applying this 3$\sigma$
rejection, we give a summary of the GC system properties for 
each galaxy in Table 2. The number of GCs per galaxy ranges 
from 39 (NGC 1700) to 328 (NGC 4365). 
For the sample as a whole the mean color is V--I = 1.09. 
Assuming that V--I colors correspond to metallicity, we also include
in Table 2 the inferred metallicity using the relation from
Couture \etal (1990), i.e.\\

[Fe/H] = 5.051~(V--I) -- 6.096 \hspace{9cm} (4)\\
%(V--I) = 0.198~[$Fe/H$] + 1.207\\

Next we examine the projected radial
distribution of GC magnitude, color and surface density.  
Globular cluster magnitude as a function of galactocentric radius is 
plotted in Fig. 7. 
Most galaxies are consistent with no mean GC magnitude variation with
distance from the galaxy center. We note that there are slight differences
for some galaxies, for example 
NGC 4494 appears to have a small enhancement of bright GCs in the
central kpc whereas for NGC 5982 they seem to be lacking.    

In Fig. 8 we show the radial distribution of GC color. 
We have fit a linear regression line over the full range of 
galactocentric radius for each galaxy. An iterative fit 
was performed, with each data point inversely weighted by its error
from {\it daophot} (which gives less weight to fainter
GCs). The slopes are listed in Table 2. 
Most of the sample galaxies show little, or no, evidence for a GC color gradient with 
galactocentric radius. There are six galaxies which show weak
radial color gradients.  
We also show in Fig. 8 the color gradient of the
underlying galaxy at small radii from Carollo \etal (1996). Although
the GC colors overlap with that of the galaxy starlight, the GC colors
are systematically bluer on average than the galaxy itself. The
difference for the whole sample is $\Delta$ (V--I) $\sim$ 0.10
$\pm$ 0.06. 

We show the surface density, per kpc$^2$, of GCs with galactocentric 
radius in Fig. 9. The GC counts have been 
restricted to 
position angles covering 
only one hemisphere, or side, of the galaxy (i.e. WFPC2 angles of 
V3 $\pm$ 90$^{\circ}$). For this hemisphere, the WFPC2 field-of-view
provides almost complete coverage (we are missing $<$1\% of the area) 
out to some fixed galactocentric distance (i.e. 100$^{''}$). 
To estimate the total surface density we multiply the number of GCs 
by two. Fig. 9 shows the
average surface density in several annular bins with Poisson errors,
and the underlying galaxy light profile normalized by an arbitrary
vertical offset.

The GC surface density profile for all galaxies appears to rise less
steeply, than the galaxy light, towards the galaxy center. 
For two galaxies
(NGC 5813 and IC 1459) there is even an indication that the number
density falls in the central kpc. 
In order to quantify the profiles we
have fit a `core model' of the form $\rho = \rho_o ($r$_c ^2 +
$r$^2)^{-1}$. 
The measured core radii (r$_c$) are given in Table 2 and plotted  
against galaxy absolute
magnitude in Fig. 10. 
A weighted fit gives r$_c$ = --0.62$\pm$0.1~M$_V$ -- 11.0. 
A correlation is seen in the sense that more luminous
galaxies have a larger GC system core radii. The correlation
between GC core radius and galaxy effective radius is weaker than for
galaxy luminosity. 

As noted above, we have almost complete areal coverage of
180$^{\circ}$ allowing us to also examine the azimuthal distribution
of GCs on one side of the galaxy.  
Fig. 11 shows the azimuthal distribution for the sample 
galaxies $\pm$90$^{\circ}$ from the galaxy major axis. 
Some galaxies
show high counts close to 0$^{\circ}$ indicating that the GC system is
closely aligned with the galaxy major axis. 

\section{Discussion}

\subsection{Comparison with ground--based data}

Three of the galaxies in our sample have had their GC systems studied
previously using ground--based data. 
The single filter photographic studies of NGC 4278 (Harris \&
van den Bergh 1981), NGC 4406 (Hanes 1977; Cohen 1988) 
and the CCD study of NGC 4365
(Harris \etal 1991) generally concentrated on GCLFs and their
usefulness as distance indicators. 
Harris (1991)
quotes a specific frequency (S$_N$) for these galaxies of 6.1 for NGC
4278, 7.7 for NGC 4365 and 5.4 for NGC 4406 (Harris assumed h =
0.75). The specific frequency is a measure of the relative richness of
a GC system 
normalized to the parent galaxy luminosity. 
%S$_N$ = N$_T 10^{0.4(M_{V} + 5 log h + 15)}$\\
Multicolor CCD data are available
for NGC 4365 and NGC 4406 from Ajhar \etal (1994). These data were
obtained on the KPNO 4m as part of a surface brightness fluctuation
study and have a magnitude limit of about V $<$ 22.5.  
We have applied the same 3$\sigma$ rejection
criterion to their GC data (kindly supplied by E. Ajhar) and find a 
mean color of V--I = 1.10 $\pm$ 0.01 for 125 
GCs in NGC 4365 and V--I = 0.98 $\pm$ 0.01 for 96 GCs in NGC 4406. 
These colors are in very good agreement with
those quoted in Table 2. Although their sample is smaller and has a
brighter cutoff magnitude, it extends to slightly larger 
galactocentric distances. We have combined our sample with that of
Ajhar \etal to show the radial variation of GC color for NGC 4365 and
NGC 4406 (see Fig. 12). 
Ajhar \etal concluded that there
was little, or no, radial gradient in either NGC 4365 or NGC 4406 GC 
colors. Our
data extend the range to smaller radii and strengthen this conclusion. 
%Ajhar \etal did not discuss the
%variation of magnitude with galactocentric radius, the number density
%profile or azimuthal distribution of these globular clusters. 

\subsection{Magnitude Distribution}

Globular cluster luminosity functions (GCLFs) can help to constrain GC
formation theories and have been used as standard candles for distance
determinations. Their use as a distance indicator is somewhat 
controversial as it 
depends on the universality of the absolute magnitude of the turnover of
the GCLF (see Harris 1991). It has been suggested that the turnover 
depends on metallicity (Ashman \etal 1995) and the dispersion of the
GCLF itself (Secker \& Harris 1993). These issues, and an estimate of the
Hubble constant from our GCLFs, will be considered in a future paper and
are not discussed further here. 

The variation of GC properties with galactocentric radius is of 
interest for formation and evolutionary theories of GCs. 
Although our data cover only a limited radial range (as set by the WFPC2
field-of-view), the central galaxy regions might be expected to show the most
obvious signatures of some formation and destruction processes. For
example, GC formation from two merging spirals, may be
expected to preferentially form new GCs in the inner regions (Ashman
\& Zepf 1992). 
%The formation from primordial giant molecular clouds suggested by
%Harris \& Pudritz (1994) predicts a uniform mass distribution with
%galactocentric radius. 
The destruction processes, such as dynamical
friction and tidal stripping are negligible beyond $\sim$3 kpc
(provided GCs are not on highly radial orbits), so that GCs
at these distances reflect their formation properties (Murray \& Lin
1993).  Interior to this, low mass GCs may be subject to dynamical
destruction processes (e.g. Aguilar, Hut \& Ostriker 1988). 

The most notable feature of Fig. 7 is that, for most galaxies, GC
magnitude (and probably mass since the variation in M/L is expected to be small) 
does not vary with distance from the
galaxy center. This uniformity is consistent with the formation
scenarios of Harris \& Pudritz (1994) and Vietri \& Pesce (1995). 
Although there is some evidence
for a radial trend in our Galaxy (van den Bergh 1995a). 
As noted in section 3.2, small differences are
seen for some galaxies. 

\subsection{Color Distribution}

The color distributions for the whole sample between 
0 $<$ V--I $<$ 2 
are shown in Fig. 6 (which have been corrected for Galactic extinction
but not internal reddening). 
The most obvious impression from
Fig. 6 is that the average globular cluster color 
does not vary much from
galaxy to galaxy. As discussed in section 4.5, this is partly an effect
of the limited range in galaxy luminosity for our sample. 
Most GCs have colors between 0.5 $<$ V--I $<$ 1.5. 
Although the mean color does not
depend on magnitude, the dispersion in color 
increases significantly at fainter magnitudes. Is the spread in GC
colors intrinsic to the GC system (suggesting a range in GC
metallicities) or is it simply due to photometric errors ? We have
compared the distribution of colors for the simulated GCs to those of
the real GCs in NGC 4365 (a galaxy with no obvious dust and the
richest GC system in our sample). We find that the dispersion in NGC
4365 GC colors is greater, particularly at bright magnitudes, than
expected from simulations. Thus although photometric errors play a
role, there is some evidence that GC systems reveal an intrinsic range
in color, and hence metallicity, within a single galaxy. This is also
the situation from previous photometric and 
spectroscopic studies (see review by Harris 1991). 
Some galaxies have broad (NGC 4589 and NGC 5813) 
or possible bimodal color distributions (NGC 4494 and IC 1459). 

We note that Ajhar \etal (1994) found a wide range of color
distributions for their 10 ellipticals. This is probably partly due to
contamination by foreground stars and small background galaxies in
their ground--based data. After applying a 3$\sigma$ color selection to
their data, we find that the galaxies with over 25 GCs have a mean
V--I color and
dispersion of 1.07$\pm$0.04 (NGC 3379), 1.10$\pm$0.01 (NGC 4365), 
0.92$\pm$0.02 (NGC 4374), 0.98$\pm$0.01 (NGC 4406), 1.08$\pm$0.01 (NGC
4472) and 0.99$\pm$0.02 (NGC 4552). So after excluding objects
with extreme colors, the Ajhar \etal data set gives a similar result
to ours, namely a fairly uniform color distribution between GC
systems. 

Bimodal, or multimodal, color distributions are expected in the merger
model for globular cluster formation of Ashman \& Zepf
(1992). During the early stages of the merger a 
young, and initially blue, metal--rich GC population is formed. 
As this population ages, it rapidly reddens and after about one Gyr, can be 
characterised as a relatively metal--rich and red population of GCs. 
A bimodal metallicity distribution is seen in our Galaxy, with the
metal--poor population associated with the halo and the more
metal--rich with the disk. In this case the bimodality, 
may or may not, have been the result of a merger. 
To date, the few detections of bimodal distributions in
ground--based data of early type galaxies have 
been somewhat controversial. Zepf \& Ashman (1993) note that unimodal
distributions can be ruled out in the Fornax cD galaxy NGC 1399, 
the dust lane elliptical NGC 5128 and 
the giant Virgo elliptical NGC 4472. 
In the case of NGC 4472, Ajhar \etal (1994) obtained
a larger number of GCs compared to that available to Zepf \& Ashman 
(1993), for which they described the V--I color distribution as
``uniform'' between 0.85 and 1.30. Furthermore, 
Ajhar \etal found no visual evidence for
bimodality in any of their other 9 early type galaxies, although they
did not test for bimodality in a statistical way. Most recently, Geisler,
Lee \& Kim (1996) present a still larger sample of GCs in NGC 4472, in
which they report strong bimodality. 

Elson \& Santiago (1996) have presented WFPC2 data on GCs 
that lie $\sim$ 2.5 arcmin from the center of M87. Their V--I color
distribution reveals a striking bimodality contrary to that of
previous ground--based data (Lee \& Geisler 1993; Couture, Harris \&
Allwright 1990). 
In this case the 
bimodality is supported by 
Whitmore \etal (1995) who studied GCs in the
center of M87 with WFPC2. 
Neither Elson \& Santiago nor Whitmore \etal discuss the radial
distribution of their color subsamples.

The histograms of GC colors for our sample (Fig. 6) reveal a fairly
symmetric distribution about a mean value, with two possible exceptions
being NGC 4494 and IC 1459. In order to quantify the color
distributions, we have carried out a
Hermite analysis, to V = 24, 
for all galaxies. This tests the distribution for non--Gaussian
shapes. In no case do we find a statistically significant
(i.e. $>$3$\sigma$) deviation from a Gaussian profile. So although the
color distributions for NGC
4494 and IC 1459 visually appear to be non--symmetric, the effect is not
statistically significant given our sample size and errors. This does
not rule out the possibility that with larger samples, deviations from
a symmetric single--peaked Gaussian will be found.  
%Color--magnitude diagrams for our sample (Fig. 4) show
%that there is very little variation of color as we go to fainter
%magnitudes for all galaxies. 
The two visual 
peaks are at V--I $\sim$ 0.95 and 1.15 for NGC 4494, and V--I $\sim$
1.05 and 1.25 for IC 1459. Compared to the sample mean of $<$ V--I $>$
= 1.09, the peaks in NGC 4494 straddle the mean, whereas 
for IC 1459 one peak is close to the mean value and the other appears
to be an `additional' red population. 
In order to investigate the spatial location of these color
subsamples, we have examined their distribution in polar coordinates,
i.e. as a function of position angle and radius. 
We have defined subsamples of 0.5 $<$
V--I $<$ 0.95 and 1.15 $<$ V--I $<$ 1.6 for NGC 4494, 
and 0.5 $<$ V--I $<$ 1.05 and 1.25 $<$ V--I $<$ 1.8 for IC 1459. 
%For NGC 4494 is there a slight enhancement of
%blue GCs with position angle $\sim$ --50$^{\circ}$ that is not seen in
%the red sample (see Fig. 11). For IC 1459 there are several red GCs with 
%position angle $\sim$ 100$^{\circ}$ (see Fig. 12). 
For both galaxies there is no strong 
trend with position angle nor galactocentric radius.
(i.e. the color subsamples are not 
centrally concentrated). 
To summarize, the current situation is that there is good evidence for
bimodality in a small number of ellipticals but many 
ellipticals show little or no evidence for bimodality. 
This may simply be due to the
level of photometric errors or a limited number of GCs in each sample and
remains an outstanding issue. 
%For
%comparison, the mean color for the blue objects in NGC 1275 is
%V--I $\sim$ 0.6 (Holtzman \etal 1992), for NGC 7252 it is $\sim$ 0.7
%(Whitmore \etal 1993) and for NGC 4038/4039 (Whitmore \& Schweizer 
%1995), after correcting for dust reddening, V--I $\sim$ 0.5.

The results from ground--based data on radial 
gradients in GC properties has been summarized by Harris (1993). He
concluded that for the Milky Way, M31, M104, NGC 5128 and M87 there is little
evidence of color gradients for galactocentric distances beyond 4
kpc. However, definite gradients did exist for NGC 4472, NGC 4649 and
NGC 1399 between 3 and 10 kpc. 
Our data probe, for the first time, regions less than a
kpc from the galaxy center, and out to several kpc.  
Our sample galaxies have weak or non--existent GC color gradients
(see Fig. 8). 
Five galaxies show a slight reddening trend close to the galaxy center
which could in part be due to dust at small radii.
One galaxy, 
NGC 4589, suggests a blueing trend, but this is probably due to dust
that is distributed throughout the galaxy even at large radii. 
Fig. 13 shows a plot of
V--I color vs radius for the whole sample. Again only a weak, 
but not statistically significant, reddening trend is seen.

\subsection{Evidence for Young Globular Clusters ?}

HST observations 
of NGC 1275 (Holtzman \etal 1992), NGC 7252 (Whitmore
\etal 1993) and NGC 4038/4039 (Whitmore \& Schweizer 1995),  
which are thought to have recently undergone a merger,
reveal a population of bright (V $\sim$ 21), blue 
(V--I $\sim$ 0.6) semi--stellar objects. These blue objects tend to be centrally
located and larger (by a factor of 2--3) than the red
objects. Such characteristics suggest that they are young
GCs formed in the merger of two gas--rich galaxies. On the other hand,
van den Bergh (1995b) has argued that these objects are more like {\it
open} rather than {\it globular} clusters. 

Examination of Fig. 5 in Whitmore \etal
(1993) clearly reveals several bright, blue objects with many fewer
red objects. Given the presence of dynamically--young dust and 
gas features in our sample galaxies, do 
we see evidence for a population of bright, blue and centrally
located GCs ? A quick visual inspection of the color
maps in Carollo \etal (1996) shows straight away that we do not find a
large population of such GCs. Ruling out a smaller population of
faint, blue GCs in our sample is difficult due to the dust. As
indicated in section 2.1, for non--dusty galaxies such as NGC 4365, we
detect the vast majority ($\sim$ 95\%) of sources in the PC
frame. Even for 
the galaxies with more extensive dust (NGC 4278, NGC 4589 and IC
1459), the fraction of totally obscured GCs is probably not large
given the relatively small covering fraction of dust.

We have visually inspected the PC image of each galaxy, searching by
eye for any possible GCs not detected by {\it daophot}, particularly
those along the edges of dust lanes. 
Although not objective, the eye is remarkably good at pattern
recognition and detects $\sim$5 faint, pointlike objects per galaxy
that were not detected by {\it daophot}.  
These objects tend to be much more elongated than the detected GCs and
have magnitudes 23 $<$ V $<$ 25. They have a wide range of colors from
very red to very blue, with an average color similar to that of the brighter
detected GCs. There is no evidence for an extended blue tail in the
color distribution of very faint objects. Indeed their extreme colors
can largely be explained by photometric errors at such magnitudes. 

Thus although there may exist a handful of
genuine GCs with blue colors, they are not a dominant
population at any magnitude or spatial location. Follow--up
spectroscopy with 8--10m class telescopes  
may be the only way to determine unambiguously whether
the blue objects are young globular clusters. 

\subsection{Metallicity Distribution}

In so far as V--I color corresponds to metallicity (i.e ignoring age
effects), we have used the
relation of Couture \etal (1990), to give the range
in GC metallicity listed in Table 2. The mean metallicity of the whole
sample is [Fe/H] = --0.62. The most metal poor ([Fe/H] = --1.10) GC
system is that of NGC 4406. The most metal rich is that of NGC 4589,
but the GC colors are probably reddened by the extensive dust present
throughout the galaxy. For comparison, Milky Way GCs peak at 
[Fe/H] $\sim$ --1.6 for halo GCs and --0.5 for disk GCs, with the ratio
of halo to disk GCs being about 5:1. Thus our elliptical galaxy 
GC systems have
metallicities similar to those of the Milky Way disk GCs and not
halo GCs. 
The weak radial color gradients seen in Fig. 8 
suggest also that the radial metallicity gradients are quite small, with
the less metal rich GCs at large distances. 
With a larger radial coverage, more convincing trends are seen in other
galaxies, e.g. M87 (Lee \& Geisler 1993). A strong metallicity gradient
is indicative of dissipational collapse, although a 
gradient in the same sense is also 
expected in the Ashman \& Zepf (1992) merger model. 

In Fig. 8, we show the variation of V--I color for the GCs and the 
underlying galaxy in the central kpc from Carollo \etal (1996). The
mean systematic offset of $\Delta$ (V--I) $\sim$ 0.10 $\pm$ 0.02 
corresponds to a
metallicity offset of $\Delta$ [Fe/H] $\sim$ 0.50 dex. 
This is in the sense that GCs are
more metal poor than their parent galaxies at any given galactocentric
distance. For the few GC systems studied to date, spectral metallicity
determinations also give $\Delta$ [Fe/H] $\sim$ 0.5 dex between the GCs and the
parent galaxy (Brodie \& Huchra 1991; Harris 1991). Thus the photometric and
spectroscopically derived metallicity determinations give similar
values. These differences 
are consistent with a scenario in which 
GCs formed before the stars in their parent galaxy (though one can not
rule out subsequent formation from low metallicity gas). 

It has been suggested that the mean metallicity of a GC system
increases with the parent galaxy luminosity for all galaxy types 
(e.g. van den Bergh 1975; Brodie \& Huchra 1991). 
This relation has been questioned by Ashman \& Bird (1993) who
suggest that no such correlation exists for non--spirals (they
excluded ellipticals from their analysis), and that all halo GC 
systems have a mean metallicity of [Fe/H] $\sim$ --1.6 irrespective of
the parent galaxy luminosity. However, Perelmuter (1995) found 
that the dwarfs, spirals and irregular galaxies {\it do} show a trend of
metallicity with luminosity. 
Ellipticals are offset from this relation to
higher mean metallicity for a given luminosity. This
offset is interpreted by Perelmuter as 
a consequence of elliptical galaxies having formed by the
merger of two spirals. The exact nature of the GC metallicity --
galaxy luminosity relation has important implications for galaxy and
GC formation and these contradictory findings need to be resolved. 
A crucial ingredient in resolving this issue is to have
a large sample of galaxies, particularly at intermediate luminosities
in order to explore the parameter space between dwarfs and massive
galaxies. Of the previous studies, Brodie \& Huchra (1991) had a
sample of 10 galaxies, Ashman \& Bird (1993) used 7 and Perelmuter
(1995) had 17 galaxies. None of these studies had many galaxies with
luminosities --17 $<$ M$_V$ $<$ --21.
% to directly test the idea that
% there was a continuity between the dwarf and massive galaxies. 

To further investigate this issue, we  
have combined the spectroscopic compilation of Perelmuter (1995)
with metallicities, inferred from V--I colors,  
for our sample and that of Ajhar \etal (1994). 
This gives by far the largest total sample
to date of 32 galaxies, with many at intermediate luminosities. 
The GC mean metallicity is plotted against parent galaxy luminosity (h
= 0.75) 
for this combined sample in Fig. 14. Data for
NGC 4472 are plotted twice, giving both spectroscopic ([Fe/H] =
--0.80) 
and photometric ([Fe/H] = --0.65) metallicities.   
Although there may be some offset in
the photometrically versus spectroscopically determined metallicities, 
it is a relatively small effect. 
An important distinction from some earlier studies, is that we have 
have subtracted off a disk component for the spirals 
using the mean bulge-to-disk ratios for
different Hubble types given by Simien \& de Vaucouleurs (1986). This
gives an approximate bulge magnitude for each spiral galaxy which is
probably more appropriate to use than the total galaxy luminosity, as
most of our Galaxy's GCs are associated with the bulge. 
This is not practical for the SMC (M$_V$ = 
--16.9) and the LMC (M$_V$ = --18.6) for which we have used total
luminosity.  
We have included new galaxies from Ajhar \etal
(1994) that have more than 25 GCs after 3$\sigma$ color rejection
(i.e. NGC 3379, NGC 4374 and NGC 4552) but 
excluded NGC 4589 as its GC colors are probably strongly affected
by dust reddening. 
A weighted linear regression fit to the data points 
gives:\\

[Fe/H] = --0.17$\pm$0.01~M$_V$ -- 4.3 \hspace{9.5cm} (5)\\

\noindent
We have not made any effort to ensure that the error bars are
consistent between the different data sets, nor to correct for any
systematic offset between data sets, and so we can not 
speculate about whether the scatter in the relation is intrinsic or
not (however for the luminous galaxies the rms scatter 
appears to be $\sim$ 0.4 dex in metallicity).   
However, we note that an 
unweighted fit 
%(which may be more appropriate given the range in error bars) 
gives a similar relation, i.e. [Fe/H] = --0.16$\pm$0.02~M$_V$ -- 4.1. 
For the fits we have used only the spectroscopically determined [Fe/H]
for NGC 4472. Brodie \& Huchra (1991) found a slope of --0.13 $\pm$
0.02 for a
small sample of GC systems. The galaxies themselves obey a
similar metallicity--luminosity relation, and for a large sample ($\sim$ 70),
Brodie \& Huchra (1991) found the slope to be --0.17 $\pm$ 0.03. 
Thus with a
larger number of GC systems, we find that the GC
metallicity--luminosity relation has the same slope as the galaxy
metallicity--luminosity relation, with a form of Z $\propto$
L$^{0.4}$. This suggests that the same physical 
process governing gas enrichment occurred in both GCs and galaxies. 

Da Costa \& Armandroff (1995) have identified 3 or 4 GCs associated
with the recently discovered Sagittarius dwarf spheroidal galaxy
(Ibata, Gilmore \& Irwin 1994). The 4 GCs give an average metallicity
of [Fe/H] = --1.4, and if we exclude the GC with a peculiar metallicity, 
the remaining 3 give [Fe/H] = --1.8. The 
correlation in Fig. 14 would predict M$_V$ = --17.0 to --14.7 for this 
range of metallicity. The
absolute magnitude of the Sagittarius dwarf is 
uncertain, Da Costa \& Armandroff suggest that it is around M$_V$
$\sim$ --15. So the GC system in the Sagittarius dwarf also appears to
be consistent with the metallicity--luminosity correlation. 

Ajhar \etal (1994) claim that there is a dependence of GC metallicity
on environment in their data set. They point out that both NGC 4374
and NGC 4406, which lie close to the center of the 
Virgo cluster potential well, have relatively blue (i.e. low
metallicity) 
globular clusters. Indeed both of these galaxies lie
marginally below
the locus of galaxies that have a similar luminosity in Fig. 14, but given
the uncertainty in both photometrically and spectroscopically derived metallicities
it would be premature to draw any conclusions at this stage. 

To summarize, we find a correlation between 
GC system mean metallicity and parent galaxy luminosity (i.e. $Z
\propto L^{0.4}$)  
over almost 10 magnitudes, confirming the suggestion first made by 
van den Bergh (1975). 
The continuity appears to extend from dwarf ellipticals through the
Magellanic Clouds and spiral bulges to giant ellipticals, unlike that
seen in the smaller samples of 
Ashman \& Bird (1993; there is no
relation for dwarfs and spirals) 
and Perelmuter (1995; the relation exists only for spirals). 
This suggests a
common origin for the GC populations over a large range of galaxy
types.
Although some type of metallicity--luminosity relationship would be
expected in the merger of two spirals, it needs to be shown that such a
merger will produce the type of relationship shown in
Fig. 14.

The work of Guzm\'an \etal (1993) on the fundamental plane of
spheroidal systems suggests that the only difference between dwarf and
giant ellipticals is the relative scaling between the size of the
initial dark halo and the size of the final light distribution. In
particular, dwarf ellipticals obey the same luminosity--sigma and
metallicity--sigma relations as giant ellipticals. Thus qualitatively, 
we can explain the GC mean metallicity vs galaxy luminosity relation
in terms of depth of the potential well and the galaxy's ability to
retain metal rich gas. 
The continuity in this relation from dwarf to giant 
ellipticals fits in well with the continuity seen in GC specific
frequencies and the galaxy
fundamental plane relations. 

Although we can not rule out alternatives at this stage, 
the similarity of the metallicity--luminosity relation
for GCs and the galaxy stellar population, and the small offset to
lower metallicity seen in GCs seems consistent with the idea that 
the majority of GCs in elliptical galaxies formed during a coalescence
phase, perhaps shortly before the main epochs of star formation 
(see also Harris 1993). 

\subsection{Surface Density Distribution}

Surface density profiles for our sample are shown in Fig. 9. This
figure indicates that the number of GCs per unit area rises less
steeply than the underlying galaxy, i.e. it does not continue to rise
as a power law.
Such effects are difficult to see from ground--based
data but have been seen  in a few cases (see Harris 1991).
Other HST studies also find this effect 
(e.g. Grillmair \etal 1994a). 

All fourteen of our
sample galaxies reveal a turnover, or core, in their log surface density
distribution. There are several effects which may give the appearance
of a reduced rise in GC density in the inner regions: 
1) a brighter limiting magnitude cutoff in the PC image,   
2) changes in the GC luminosity function with radius, 
3) obscuration due to dust, 4) crowding of GCs. 
The first two possibilities do not appear
to be significant since our simulations showed no radial variation in 
magnitude, and the GC luminosity functions are consistent within sampling
errors for GCs at small and large galactocentric
distances. Furthermore the detection fraction is similar from
CCD-to-CCD, so that the detection characteristics in the PC image are
similar to those in the WFC images. 
The third possibility can not be ruled out, but the galaxies with no
obvious dust (NGC 4365 and NGC 5982) also have clear 
cores in their surface density distributions. Finally, 
even in the PC image, the density of GCs is not so high that they are
blended together. So crowding can be ruled out. 
Thus the cores seem to be real, and 
typically have sizes of r$_c$ $\sim$ 2 kpc. 
In Fig. 10 we show a weak correlation between core radius 
and galaxy luminosity. This correlation is in the sense that 
more massive ellipticals have larger core radii. Alternatively, 
the most luminous galaxies (i.e. those
with M$_V$ $\le$ --21) have large cores, but for the small, rotating ellipticals the
core radius is weakly, if at all, dependent on galaxy luminosity. 

The Milky Way galaxy also
shows a reduction in the surface density of GCs interior to about 1
kpc, and if we used the magnitude of the bulge it would lie close to
the fit in Fig. 10. 
The core in the Milky Way GC distribution 
has been attributed to destruction processes that have occurred 
after formation
(e.g. Aguilar \etal 1988). However, in the case of NGC 3311,
Grillmair \etal (1994a) concluded that the core radius of the
distribution was too large to have been caused by dynamical friction
even if operating for a Hubble time. Similar arguments can be made for
the galaxies in our sample with the largest core radii. Furthermore,
we do not expect dynamical friction, or tidal shocking, to produce a
core radius with a steep dependency on galaxy mass as suggested by
Fig. 10. As destruction processes seem unlikely to explain the core
distribution, the simplist explanation may be that the GC 
distribution is established at the epoch of formation.

\subsection{Azimuthal Distribution}

In Fig. 11 we show the distribution of GCs as a function of position
angle about the major axis. 
%A spherically symmetric GC distribution  would
%be represented by a uniform position angle distribution. 
In several cases the GC distribution peaks close to the 
galaxy major axis. There are also notable exceptions, e.g. both NGC
4589 and IC 1459 have peaks close to 45$^{\circ}$ to
the major and minor axes. For NGC 4406, there is a small enhancement
at position angle $\sim -75^{\circ}$ due to 
a companion (?) galaxy with its own small GC population. 
Combining the GC systems of all galaxies, we
find a mean position angle that is consistent with the major axis
of the galaxy itself (i.e. 3 $\pm$ 15$^{\circ}$). 
This is also the case if we restrict the galaxies more elongated than
E1 for which the position angle is better defined. 
The mean ellipticity
is given by the quadrature sum of 
$<sin 2 \phi>$ and $<cos 2 \phi>$ for a radial density distribution of
$\rho \propto r^{-2}$. This gives an average ellipticity for the GC
systems of E0--E1, i.e. slightly rounder than the average galaxy light
(see Table 1). 
%Such a test would be
%better suited to a highly elliptical galaxy for which 360$^{\circ}$
%coverage was available out to large radii. 
The azimuthal distribution in a few other galaxies
have been studied, e.g. M87 (McLaughlin \etal 1994).
McLaughlin \etal find that the GC system of M87 has a major axis that
roughly coincides with the galaxy isophotes, and has an ellipticity
that increases with galactocentric distance, i.e. it is rounder near
the galaxy center and flattens in the outer regions. 

\subsection{Implications for Globular Cluster Formation}

In the last few years quantitative information about extragalactic GC
systems has grown rapidly. Such information now provides useful
constraints on possible GC formation scenarios (Harris 1993). For
example, correlations of GC properties with the parent galaxy would
argue against a pre--galactic origin for GCs, where GCs formed
before the dark matter potential wells had assembled (see Peebles \&
Dicke 1968). Futhermore, the 
properties of GCs
(e.g. kinematics, specific frequency and metallicity) do not match
those expected if they formed from a cooling flow (e.g. Harris,
Pritchet \& McClure 1995; Grillmair \etal 1994b). 
In this section we focus on the
implications of our results, and those in the literature, for two
general formation mechanisms. The first possibility is that GCs form
at roughly the same epoch as the parent elliptical during the early 
collapse and coalescence 
at high redshift (e.g. Searle \& Zinn 1978; Harris \& Pudritz
1994). Alternatively, the late merger of two spiral galaxies
gives rise to an extended population of metal poor GCs (Hernquist \&
Bolte 1993) and a centrally concentrated population of new metal rich
GCs (Ashman \& Zepf 1992). We note that studies of nearby galaxies may
only  provide circumstantial evidence on the general problem of galaxy
formation. 

Starting with the latter, Zepf \& Ashman (1993) have noted that their
model can explain many of the observational constraints. In
particular, they account for the shallower slope of the 
density profiles of GC
systems as the `puffing--up' of the metal poor GC distribution. 
This also gives rise to a metallicity (and color) radial gradients, 
that is offset to lower metallicity compared to the underlying galaxy,
at any given radius. 
Perhaps the clearest prediction
of the Ashman \& Zepf model is a bimodal metallicity (and hence color)
distribution as discussed in section 4.3.  
The new metal rich GCs are expected to be more centrally
concentrated and at least as numerous as 
those of the original metal poor GCs. Within the limitations of each
GC system size 
(the richest system contained 328 GCs), we found no
statistically convincing cases for a bimodal color distribution. 
Furthermore, the relative numbers of metal rich to metal poor GCs
suggested by those distributions is $\sim$ 1:1 and not 3:1 as might be 
needed to explain the large numbers of GCs seen in some ellipticals 
(see also van den Bergh 1995c). The number of ellipticals with clear
bimodal metallicity distributions in the literature is small and they
tend to be massive ellipticals (too massive to be formed from two
typical spirals). 
%remains to be seen if this is a common feature. 
%For
%the two galaxies in our sample 
%with possible bimodal color distributions, 
In our sample, we do not find evidence 
for a distinct red (metal rich) population that is centrally
concentrated. 
Although a population of young, blue GCs appear to have formed in 
several merging galaxies, there is little evidence
to support the presence of such objects in present day ellipticals, 
even in those with unsettled dust distributions indicative of a 
recent accretion event.
Perhaps the most interesting new results of this work, that need to be
understood in the context of the Ashman \& Zepf (1992) merger
model, are the correlations of GC surface density core radius 
and GC mean
metallicity with parent galaxy luminosity.

Similarly, no firm predictions exist for scenarios in which GCs form
in an early collpase of unevolved galaxies, as most simulations of galaxy
formation can not yet resolve details at this level. The discussion
therefore remains somewhat schematic, and can be only tested for rough
consistency. 
%In terms of an early collapse scenario for GC and
%galaxy formation, 
With these caveats, the differences between the 
GC density and metallicity
profiles to those of the underlying galaxy can be understood if the GC system
formed slightly earlier in the initial coalescence phase than the galaxy
itself (see Harris 1993). Thus for a given galaxy, 
GCs form in a more extended proto--galaxy than the galaxy's stars, 
resulting in a lower metallicity and more extended distribution for
the GCs. 
For a virialized system, 
the variation in these properties is correlated with the depth of the
potential well, which is in turn presumably related to the galaxy total
luminosity. This scenario is similar to that envisaged by Eggen,
Lynden--Bell \& Sandage (1962) and would suggest that the difference in
metallicity at a given galactocentric radius (i.e. $\Delta$~[Fe/H]
$\sim$ 0.5 dex) corresponds to the timescale for galaxy collapse and 
star formation. 

\section{Summary and Conclusions}

Very little is known about globular clusters in the central regions of
elliptical galaxies, due to the effects of seeing/crowding on a bright
galaxy background, in current ground--based observations. These problems are
largely overcome using {\it HST}, which has the additional advantage
that the effective 
contamination from foreground stars and background galaxies is
very small. From WFPC2 imaging, we have
identified globular clusters in the central few kpc of 14 elliptical
galaxies with kinematically--distinct cores (generally 
thought to be the result of an accretion of a gas--rich galaxy)
These regions, close to the galaxy center, are where we expect to see
the strongest signatures of some formation and destruction processes. 
Our sample substantially 
increases the number of extragalactic 
globular clusters systems studied to date. 

After subtracting the background galaxy, we have used {\it daophot} to
detect globular cluster candidates. 
We then measure sizes and 
magnitudes for these objects. Objects close to hot pixels, or those with
extreme colors or large sizes, are removed from the candidate list
giving $\sim$ 200 objects per galaxy, the vast majority of which are
globular clusters. We have determined accurate positions, V
magnitudes, V--I colors and the spatial distribution of these globular
clusters to a limiting magnitude of V $\sim$ 25. The globular cluster
sample is complete to V $\sim$ 24. Comparison with the V
$\le$ 22.5 KPNO 4m observations of Ajhar \etal (1994) show good
agreement for the two galaxies in common. 

Before examining the magnitude, color and spatial distribution of our
sample globular clusters, we performed simulations to ensure that any
trends seen were real and not simply some selection effect. 
In confirmation with previous ground--based studies, we find little or
no trend for the globular cluster luminosity (and therefore mass, since 
M/L changes are expected to be small) 
to vary with distance from the galaxy center
(galactocentric radius). 
Although we derive luminosity functions, we defer their discussion and
any implications for the Hubble constant to a future paper. 
We find that the mean globular cluster color is remarkably constant  
with magnitude, and that the spread in color, at any given magnitude,
is intrinsic to the GC system. 
The fact that the mean color is 
very similar from galaxy to galaxy (i.e. $<$V--I$>$ = 1.09 $\pm$ 0.01), 
probably reflects the small range
in galaxy luminosity for our sample. Some galaxies show a weak
reddening trend at small galactocentric radii, suggesting that
globular clusters in the inner region are relatively more metal rich
than those further out. This effect does {\it not} appear to be due to
dust but we can not rule it out. Many galaxies are consistent with no radial
globular cluster color gradient. For all sample galaxies, the globular
cluster colors are bluer (more metal poor) than the underlying 
galaxy starlight at any
given galactocentric radius. This difference, $\Delta$ (V--I) $\sim$
0.10 $\pm$0.02, corresponds to $\sim$ 0.5 dex in [Fe/H] which is consistent
with that found spectroscopically 
for other galaxies (Brodie \& Huchra 1991; Harris 1991). 

Perhaps one of the most significant results from this work is the
good  correlation seen between globular cluster mean metallicity and
parent galaxy luminosity. Such a relationship was suggested by van den
Bergh (1975) and Brodie \&
Huchra (1991). 
%, but which claim to have been refuted by Ashman
%\& Bird (1993). 
We have significantly increased the number of galaxies
in the relation by including photometrically
derived metallicities for our sample and those from Ajhar \etal
(1994) giving a total sample of over 30 spheroidal systems. The correlation
implies that $Z \propto L^{0.4}$ over roughly 10 magnitudes, with a
slope that is the same as that for the galaxy metallicity--luminosity
relation.  
This continuity from dwarf ellipticals, to spiral bulges
to giant ellipticals provides a valuable constraint on
galaxy and globular cluster formation. In particular, we need to 
understand how giant ellipticals and their globular cluster population
could form as the result of a merger of two typical spiral galaxies and
reproduce the observed trend over many galaxy types and masses.
We have also examined the radial and azimuthal distribution of
globular clusters. 
We find that the surface density of globular
clusters rises less steeply than the underlying galaxy light. Such an
effect is not adequately explained by different completeness levels at
small radii, dust obscuration or blending of objects. A core model
provides a good description to the radial surface density profile. 
More luminous galaxies in our sample have larger core radii, which
would seem to rule out destruction of the globular clusters by 
dynamical friction as an explanation for the observed radial distribution. 
For most galaxies, the GCs are closely aligned with the galaxy
isophotes 
but they may have a slightly rounder distribution than the galaxy
itself. 

The correlations of globular cluster surface density 
core radius and mean
metallicity with parent galaxy luminosity provide new constraints on 
models for the formation of globular clusters. 
We suggest that the properties of globular clusters are
most consistent with their formation during the  
the early collapse and coalescence phase at high redshift. 
%Subsequent mergers
%would have typically only involved 
%small galaxies having little effect on the parent galaxy
%or its globular cluster population. 

\noindent
{\bf Acknowledgments}\\
We thank J. Brodie and C. Grillmair for many useful
discussions. We also thank the referee, S. van den Bergh, for his
many suggestions, in particular noting the alternative interpretation
of figure 10. This research was funded by the HST grant GO-3551.01-91A\\

\newpage

\noindent{\bf References}

\noindent
Aguilar, L., Hut, P., \& Ostriker, J. P. 1988, ApJ, 335, 720\\
Ajhar, E. A., Blakeslee, J. P., \& Tonry, J. L. 1994, AJ, 108, 2087\\
Ashman, K. M., \& Bird, C. M. 1993, AJ, 106, 2281\\
Ashman, K. M., \& Zepf, S. E. 1992, ApJ, 384, 50\\
Ashman, K. M., Conti, A., \& Zepf, S. E. 1995, AJ, 110, 1164\\ 
Balcells, M., \& Quinn, P. J. 1990, ApJ, 361, 381\\
Baum, W. A., \etal 1995, AJ, 110, 2537\\
Bender, R., Surma, P., Dobereiner, C., \& Madejsky, R. 1989, A \& A,
217, 35\\
%Bender, R. 1990, Dynamics and Interactions of Galaxies, p. 232, ed. 
%R. Wielen, Springer-Verlag, Berlin\\
%Bender, R., Dobereiner, S., \& Mollenhof, C. 1988, A \& AS, 74, 385\\ 
Bender, R., \& Surma, P. 1992, A \& A, 258, 250\\
Bender, R., Burstein, D., \& Faber, S. M. 1992, ApJ, 399, 462\\
%Binney, J., \& Tremaine, S. 1987, Galactic Dynamics, Princeton
%University Press, Princeton\\
Brodie, J. P., \& Huchra, J. 1991, ApJ, 379, 157\\
Burrows, C., \etal 1993, Hubble Space Telescope Wide Field and
Planetary Camera 2 Instrument Handbook, STScI\\
Carollo, C. M. \etal 1996, ApJ, submitted\\
Cohen, J. 1988, AJ, 95, 682\\
Couture, J., Harris, W. E., \& Allwright, J. W. B., 1990, ApJS, 73,
671\\
Couture, J., Harris, W. E., \& Allwright, J. W. B., 1991, ApJ, 372, 97\\
%Carollo, C. M., Danziger, I. J., \& Buson, L. 1993, MNRAS, 265, 553\\
Da Costa, G. S., \& Armandroff, T. E. 1995, AJ, 109, 2533\\
%Dubinski, J., \& Carlberg, R. G., 1991, ApJ, 378, 496\\
%Djorgovski, S. J. 1985, PhD, University of California, Berkeley\\
Eggen, O. J., Lynden--Bell, D., \& Sandage, A. 1962, ApJ, 136, 748\\
Elson, R. A. W., \& Santiago, B. X. 1996, MNRAS, in press\\
Fabian, A., Nulsen, P., \& Canizares, C. 1984, Nature, 310, 733\\
%Forbes, D. A. 1991, MNRAS, 249, 779\\
%Forbes, D. A. 1994, AJ, 107, 2017\\
Forbes, D. A., \& Thomson, R. C. 1992, MNRAS, 254, 723\\
%Forbes, D. A., Reitzel, D. B., \& Williger, G. M. 1994, AJ, in press\\
Forbes, D. A., Franx, M., \& Illingworth, G. D. 1995, AJ, 109, 1988\\
Forbes, D. A., Elson, R. A. W., Phillips, A. C., 
Illingworth, G. D. \& Koo, D. C. 1994, ApJ, 437, L17\\
%Forbes, D. A., Sparks, W. B., \& Macchetto, F. D. 1990, Paired and
%Interacting Galaxies, p. 431, ed. J. W. Sulentic, W. C. Keel and C.
%M. Telesco, NASA conference publication 3098\\
%Franx, M., \& Illingworth, G. D. 1988, ApJ, 327, L55\\
%Franx, M., \& Illingworth, G. D. 1990, ApJ, 359, L41\\
%Franx, M., \& Illingworth, G. D. 1995, in preparation\\
%Franx, M., \& Illingworth, G. D., Heckman, T. M. 1989, ApJ, 344, 613\\
%Goudfrooij, P., Norgaard Nielson, H. U., Hansen, L., Jorgensen, H. E.,
%\& de Jong, T. 1990, A \& A, 228, L9\\
Geisler, D., Lee, M. G., \& Kim, E. 1996, AJ, submitted\\
Goudfrooij, P., Hansen, L., Jorgensen, H. E., \& Norgaard Nielson, H. U.  
1994, A \& AS, 105, 341\\
%Gunn, J. E. 1979, Active Galactic Nuclei, p. 213, ed.\ C.\ Hazard and
%S.\ Mitton, Cambridge University Press, Cambridge\\
Grillmair, C. 1995, personal communication\\
%Grillmair, C., Pritchet, C., \& van den Bergh, S. 1986, AJ 91, 1328\\
Grillmair, C. \etal 1994a, AJ, 108, 102\\ 
Grillmair, C. \etal 1994b, ApJ, 422, L7\\ 
Guzm\'an, R., Lucey, J. R., \& Bower, R. G. 1993, MNRAS, 265, 731\\
Hanes, D. A. 1977, Mem. RAS, 84, 45\\
Harris, H. C., Baum, W. A., Hunter, D. A., \& Kreidel, T. J. 1991, AJ,
101, 677\\ 
%Harris, W. E. 1990, PASP, 102, 966\\
Harris, W. E. 1991, ARAA, 29, 543\\
Harris, W. E. 1993, The Globular Cluster -- Galaxy Connection, p. 472,
ed. G. Smith and J. Brodie, ASP conference series, San Francisco\\
Harris, W. E. \etal 1986, AJ, 91, 822\\ 
Harris, W. E., \& Pudritz, R. E. 1994, 429, 177\\ 
Harris, W. E., \& van den Bergh, S. 1981, AJ, 86, 1627\\
Harris, W. E., Pritchet, C. J., \& McClure, R. D. 1995, ApJ, 441, 120\\
%Hau, G. K. T., \& Thomson, R. C. 1994, MNRAS, 270, L23\\
Hernquist, L., \& Barnes, J. E. 1991, Nature, 354, 210\\
Hernquist, L., \& Bolte, M. 1993, The Globular Cluster -- Galaxy Connection, p. 788,
ed. G. Smith and J. Brodie, ASP conference series, San Francisco\\
Holtzman, J., \etal 1992, AJ, 103, 691\\
Holtzman, J., \etal 1995a, PASP, 107, 156\\
Holtzman, J., \etal 1995b, PASP, 107, 1065\\
Ibata, R. A., Gilmore, G., \& Irwin, M. J. 1994, Nature, 370, 194\\
Illingworth, G. D., \& Franx, M. 1989, Dynamics of Dense Stellar
Systems, p. 13, ed. D. Merritt, Cambridge University Press, Cambridge\\
%Jaffe, W., Ford, H. C., O'Connell, R. W., v. d. Bosch, F. C., \&
%Ferrarese, L. 1994, AJ, 108, 1567\\
%Kormendy, J. 1984, ApJ, 287, 577\\
%Kormendy, J., Dressler, A., Byun, Y., Faber, S. M., Grillmair, C.,
%Lauer, T., Richstone, D., \& Tremaine, S. 1994, ESO Workshop on Dwarf
%Galaxies, ed. G. Meylan, Garching, ESO, in press\\
%Lauer, T. R. 1985, ApJS, 57, 473\\
%Lauer, T. R. 1989, PASP, 101, 445\\
%Lauer, T. R., \etal 1991, Ap J, 369, L41\\
%Lauer, T. R., \etal 1992a, AJ, 104, 552\\
%Lauer, T. R., \etal 1992b, AJ, 103, 703\\ 
%Lauer, T. R., \etal 1993, AJ, 106, 1436\\ 
Lee, M. G., \& Geisler, D. 1993, AJ, 106, 493\\
McLaughlin, D. E., Harris, W. E., \& Hanes, D. A. 1994, ApJ, 422, 486\\
%Mollenhoff, C., \& Bender, R., 1989, A \& A, 214, 61\\
%Mould, J. 1984, PASP, 96, 773\\
Murray, S. D., \& Lin, D. N. C. 1993, The Globular Cluster -- 
Galaxy Connection, p. 738,
ed. G. Smith and J. Brodie, ASP conference series, San Francisco\\
Peebles, P. J. E., \& Dicke, R. H. 1984, ApJ, 277, 470\\
Perelmuter, J. 1995, ApJ, 454, 762\\
Phillips, A. C. \etal 1995, in preparation\\
Roberts, M. S., Hogg, D. E., Bregman, J. N., Forman, W. R., \& Jones,
C. 1991, ApJS, 75, 751\\
%Sadler, E. M., \& Gerhard, O. E. 1985, MNRAS, 214, 177\\
%Sadler, E. M., Jenkins, C. R., \& Kotanyi, C. G. 1989, MNRAS, 240,
%591\\
Searle, L., \& Zinn, R. 1978, ApJ, 437, 214\\
Secker, J., \& Harris, W. E. 1993, AJ, 105, 1358\\
Schweizer, F. 1987, Nearly Normal Galaxies, p. 18, ed. S. Faber,
Springer-Verlag, New York\\ 
%Schweizer, F. 1990, Dynamics and Interactions of Galaxies, p. 60, ed. 
%R. Wielen, Springer-Verlag, Berlin\\
Simien, F., \& de Vaucouleurs, G. 1986, ApJ, 302, 564\\
%Slee, O. B., Sadler, E. M., Reynolds, J. E., \& Ekers, R. D. 1994,
%MNRAS, 269, 928\\
%Sparks, W. B., Disney, M. J., Wall, J. V., \& Rodgers, A. W. 1984,
%MNRAS, 207, 445\\
%Sparks, W. B, Wall, J. V., Thorne, D. J., Jorden, P. R., van Breda,
%I. G., Rudd, P. J., \& Jorgensen, H. E. 1985, MNRAS, 217, 87\\
%Statler, T. S. 1991, ApJ, 382, L11\\
Stetson, P. B., 1987, PASP, 99, 191\\
%Surma, P. 1992, Structure, Dynamics and Chemical Evolution of
%Elliptical Galaxies, ed. I. J. Danziger, W. W. Zeilinger and K. Kjar,
%ESO: Garching, p. 669\\
%v. d. Bosch, F. C., Ferrarese, L., Jaffe, W., Ford, H. C., \&
%O'Connell. R. W. 1994, AJ, 108, 1579\\
van den Bergh, S. 1975, ARAA, 13, 217\\
van den Bergh, S. 1990, Dynamics and Interactions of Galaxies, p. 492, ed. 
R. Wielen, Springer-Verlag, Berlin\\
van den Bergh, S. 1995a, AJ, 110, 1171\\
van den Bergh, S. 1995b, ApJ, 450, 27\\
van den Bergh, S. 1995c, AJ, 110, 2700\\
Vietri, M., \& Pesce, E. 1995, ApJ, 442, 618\\
%Wagner, S. J., Bender, R., \& Mollenhoff, C. 1988, A \& A, 195, L5\\
Whitmore, B. C. 1995, personal communication\\
Whitmore, B. C., \& Schweizer, F. 1995, AJ, 109, 960\\
Whitmore, B. C., Schweizer, F., Leitherer, C. Borne, K., \& Robert,
C. 1993, AJ, 106, 1354\\
Whitmore, B. C., Sparks, W. B., Lucas, R. A., Macchetto, F. D., \&
Biretta, J. A. 1995, ApJ, in press\\
Zepf, S. E., \& Ashman, K. M. 1993, MNRAS, 264, 611\\

\newpage
\noindent
{\bf Figure Captions}\\

\noindent
{\bf Figure 1} (Plate ***) Grey scale WFPC2 image of NGC 1427 after
galaxy subtraction. The blank region between CCDs has been removed as
this is in the pyramid shadow. Here the PC CCD is shown full size
(upper left),
even though its area coverage is only a quarter that of the WFC CCDs.
The horizontal bar in the lower right represents a WFC scale of 30
arcsec.  
The image reveals over 150 globular clusters with 
magnitudes 19 $<$ V $<$ 25. \\
%A slight enhancement of globular clusters
%along the galaxy major axis (P.A. = 80$^{\circ}$) can be seen.\\

%\noindent
%{\bf Figure 2} CCD-to-CCD detection variations. The actual detection
%fraction with magnitude is shown for the combined data of NGC 4365 and
%NGC 5982 (two dust free galaxies). The dashed line represents the
%detection fraction in each CCD, and the solid line is for all 
%four CCDs. The PC1 and WF3 have about half as many objects as 
%WF1 and WF4. The detection fraction in all four CCDs is similar to V
%= 25. \\

\noindent
{\bf Figure 2} Photometric errors from simulations. The difference
between the input magnitude of the simulated globular cluster 
and that measured by {\it daophot} is shown against 
input magnitude. A photometric
error of $\pm$0.1$^m$ is reached at V $\sim$ 24 and
I $\sim$ 22.\\

\noindent
{\bf Figure 3a} Completeness function from simulations. The fraction of
recovered globular clusters is shown against input magnitude. A fit to
the completeness function is shown by a dashed line. The
100\% completeness level is V $\sim$ 24. {\bf b)} Variation
of the output globular cluster V magnitude as a function of
galactocentric radius from simulations. {\bf c)} Radial variation of 
output V--I color (a value of 1.0 was input). 
There are no intrinsic radial gradients in magnitude or color for
simulated globular clusters.\\

\noindent
{\bf Figure 4} Color--magnitude diagrams for the sample
galaxies. Typical error bars are shown in the top left panel. Globular
cluster color is fairly constant with magnitude.\\

\noindent
{\bf Figure 5} Globular cluster luminosity functions for the sample
galaxies. The luminosity functions have not been corrected for
completeness effects. \\

\noindent
{\bf Figure 6} Histograms of globular cluster V--I colors for the
sample galaxies. 
The mean color for the combined sample is V--I = 1.09. Most galaxies
reveal a fairly tight distribution, with two (NGC 4494 and IC 1459) 
showing evidence for a possible bimodal color distribution. \\ 

\noindent
{\bf Figure 7} Globular cluster V magnitude vs galactocentric
radius. The sample is ordered into nearby, intermediate and more 
distant galaxies. For this and subsequent figures we only plot
globular clusters within 3$\sigma$ of the mean color. 
Most galaxies show no radial gradient in
the mean magnitude for the globular cluster system.\\

\noindent
{\bf Figure 8} Globular cluster V--I color vs galactocentric
radius. The sample is ordered as in Fig. 7. Galaxies reveal a weak, or
no, radial color gradient for their globular clusters.  We
also show the V--I color gradient (dashed line) for the underlying
galaxy in the central kpc from Carollo \etal (1996). In each case the
globular cluster mean color is bluer than the underlying starlight.\\

\noindent
{\bf Figure 9} Globular cluster surface density vs galactocentric
radius. The sample is ordered as in Fig. 7. Error bars represent
Poisson errors. The dashed line represents the underlying galaxy light
normalized by an arbitrary offset. 
All galaxies reveal a turnover, or core, in the GC density 
distribution. \\

\noindent
{\bf Figure 10} Core radius vs parent galaxy luminosity. 
The solid line represents a weighted fit to the sample. 
More luminous galaxies tend to have more
extended globular cluster systems.\\

\noindent
{\bf Figure 11} Histograms of globular cluster position angles for the
sample galaxies after restricting the position angles to one
hemisphere of the galaxy. 
The galaxy major axis is located at 0$^{\circ}$. \\

\noindent
{\bf Figure 12} Globular cluster V--I color vs galactocentric
radius for NGC 4365 and NGC 4406. Our sample is represented by filled
circles and the ground--based data of Ajhar \etal (1994), after a
3$\sigma$ color rejection, by open
squares. For both galaxies the two data sets are in good agreement. \\

%\noindent
%{\bf Figure 14} Polar coordinate diagram for NGC 4494. The 
%red subsample is represented by stars and the blue 
%subsample by squares. The position angle of the major axis is indicated. \\

%\noindent
%{\bf Figure 15} Polar coordinate diagram for IC 1459. 
%The red subsample is represented by stars and the blue 
%subsample by squares. The position angle of the major axis is indicated. \\

\noindent
{\bf Figure 13} Globular cluster V--I color vs galactocentric
radius for the combined sample. A weak reddening trend towards the
galaxy center is seen.\\

\noindent
{\bf Figure 14} Mean globular cluster metallicity vs absolute V
magnitude of the parent galaxy. Circles represent 
photometrically--derived [Fe/H] and 
squares spectroscopically--derived [Fe/H]. 
Our sample is given by filled
circles  and  the ground--based data of
Ajhar \etal (1994) by open circles. 
%The error bars are from Poisson number statistics. 
%have excluded NGC 4589 because of the extensive dust
%throughout the galaxy, and NGC 4468 and NGC 4489 as a mean color could
%not be defined due to the small number of GCs. 
Spectroscopic
metallicities are from the compilation of Perelmuter (1995). Although
there may exist a small offset between photometric and spectroscopic
metallicities, there is a strong correlation between the mean
metallicity of the GC system and the parent galaxy luminosity over
several orders of magnitude. A weighted fit (shown by a solid line) 
gives [Fe/H] = 0.17~M$_V$ -- 4.3. 
Realistic errors, within and between different data sets, are
difficult to assess and so it is not clear if the scatter about the
relation is intrinsic. A non--weighted fit gives a similar
relation. 
For this plot we have assumed h = 0.75. 
The galaxies include dwarf ellipticals,
the Magellanic Clouds, bulges of nearby spirals and giant ellipticals.\\

%\noindent
%{\bf Figure 18} Histogram of globular cluster position angles for
%sample galaxies more flattened than E1. 
%Here we have shifted each galaxy's major axis so as
%to lie at position angle 0$^{\circ}$. The globular cluster counts peak
%close to the major axis and decline towards the minor axis.\\

\end{document}